\documentclass[journal]{IEEEtran}
\usepackage[cmex10]{amsmath}
\usepackage{amsmath}
\usepackage{amssymb}
\usepackage{amsthm}
\usepackage{mdwmath}
\usepackage{mdwtab}
\usepackage{eqparbox}
\usepackage{autobreak}

\usepackage{algorithm}
\usepackage{graphicx}
\usepackage{textcomp}
\usepackage{subfigure}

\usepackage{amsmath,amsfonts}
\usepackage{algorithmic}
\usepackage{array}
\usepackage{textcomp}
\usepackage{stfloats}
\usepackage{url}
\usepackage{verbatim}
\usepackage{graphicx}

\usepackage{xcolor,soul,framed} 
\colorlet{shadecolor}{yellow}
\usepackage{verbatim}
\usepackage{url}
\usepackage{cite}

\usepackage{color}

\usepackage{microtype}
\usepackage{booktabs}
\usepackage{array}
\usepackage{tabularx}
\usepackage{hhline}
\usepackage{rotating}
\usepackage{multirow}
\usepackage{booktabs}
\usepackage{threeparttable}
\usepackage{makecell}
\usepackage{stfloats}

\usepackage{nomencl}

\newtheorem{lemma}{\bf Lemma}
\newtheorem{assumption}{\bf Assumption}
\newtheorem{theorem}{\bf Theorem}
\newtheorem{proposition}{\bf Proposition}
\newtheorem{remark}{\bf Remark}
\allowdisplaybreaks[1] 

\usepackage{threeparttable}

\def\BibTeX{{\rm B\kern-.05em{\sc i\kern-.025em b}\kern-.08em
    T\kern-.1667em\lower.7ex\hbox{E}\kern-.125emX}}
\usepackage{balance}

\begin{document}

\title{Robust Distribution Network Reconfiguration Using Mapping-based Column-and-Constraint Generation}
\author{Runjie Zhang, Kaiping Qu,~\IEEEmembership{Member,~IEEE}, Changhong Zhao,~\IEEEmembership{Senior Member, IEEE}, and
\\Wanjun Huang,~\IEEEmembership{Member,~IEEE}
\thanks{This work was supported in part by Guangdong Basic and Applied Basic Research Foundation under Grant 2023A1515110671, and in part by Hong Kong Research Grants Council under Grant 14207524. (Corresponding author: Wanjun Huang.)}
\thanks{R. Zhang and C. Zhao are with the Department of Information Engineering, the Chinese University of Hong Kong, New Territories, Hong Kong, China. Email: zr023@ie.cuhk.edu.hk, chzhao@ie.cuhk.edu.hk}
\thanks{K. Qu is with the College of Electrical Engineering and Automation, Fuzhou University, Fuzhou, China. Email: qu\_kaiping@163.com}
\thanks{W. Huang is with the School of Automation Science and Electrical Engineering, Beihang University, Beijing, China. Email: wjhuang1211@gmail.com}
}



\maketitle

\begin{abstract}
The integration of intermittent renewable energy sources into distribution networks introduces significant uncertainties and fluctuations, challenging their operational security, stability, and efficiency. 
This paper considers robust distribution network reconfiguration (RDNR) with renewable generator resizing, modeled as a two-stage robust optimization (RO) problem with decision-dependent uncertainty (DDU).
Our model optimizes resizing decisions as the upper bounds of renewable generator outputs, while also optimizing the network topology. 
We design a mapping-based column-and-constraint generation (C\&CG) algorithm to address the computational challenges raised by DDU. 
Sensitivity analyses further explore the impact of uncertainty set parameters on optimal solutions. 
Case studies demonstrate the effectiveness of the proposed algorithm in reducing computational complexity while ensuring solution optimality. 
\end{abstract}

\section{Introduction}
Distribution network reconfiguration (DNR) is essential for modern power systems, enhancing operational efficiency, reliability, and security \cite{baran1989network}. 
Although distribution networks have a meshed structure, they are typically operated radially by keeping certain branches open to facilitate protection \cite{lavorato2011imposing}. 
The reconfiguration process involves determining an optimal topology to minimize operational costs while meeting system security requirements.
Mathematically, DNR is expressed as a mixed-integer nonlinear optimization problem, which is challenging to solve due to quadratic power flows and binary switching variables constrained by radial topology \cite{taylor2012convex}. 
Many studies focused on the heuristic algorithms with lower computational complexity, such as iterative branch exchange \cite{civanlar1988distribution}, successive branch reduction \cite{peng2014feeder}, and switch opening and exchange \cite{zhan2020switch}, but the optimality of these algorithms cannot be guaranteed in the general DNR problem.
By leveraging the convexification techniques, a great deal of literature has recently studied DNR problems as a mixed-integer second-order cone program (SOCP), making it possible for efficient computation using commercial solvers \cite{jabr2012minimum, taylor2012convex, lee2014robust}.

With the increasing integration of renewable generators (RGs), DNR optimization has faced problems with safe operation and computational complexity due to the uncertainty of the RGs.
Addressing such uncertainty to ensure system security and maintain power balance has gained considerable attention \cite{impram2020challenges}.
There are three popular techniques to address the renewable generation uncertainty in the DNR problem: robust DNR \cite{lee2014robust}, stochastic DNR \cite{zhan2020switch}, and distributionally robust DNR \cite{zheng2020adaptive}.
The stochastic DNR aims to minimize the expected cost based on assumed known probability distributions of RG outputs \cite{ zhan2020switch }. 
However, exact probability distributions of RGs are inaccessible in practice, and the safety may be compromised due to insufficient attention to extreme scenarios. 
Distributionally robust techniques make decisions robust against uncertainty across a set of probability distributions informed by historical data, thus balancing economy and safety, but they are computationally intensive and need a large amount of data to ensure the statistical robustness. 
Therefore, we consider the robust optimization (RO) techniques, which have been widely applied to the DNR with uncertainty for the computational tractability and security of power systems \cite{lee2014robust, haghighat2015distribution}.
Robust DNR considers all possible realizations of RGs and solves for the optimal decision under the worst-case scenario, which can provide a conservative but safe solution with relatively low computational complexity. 

In prior studies, the uncertainty set in robust DNR is predefined based on historical data and advanced measurements, assuming independence from decisions or other factors. 
However, some practical strategies taken in the preparatory schedule, such as resizing the output limits of RGs \cite{ chen2022robust}, can alter the size of the uncertainty set to reduce curtailment and enhance the stability of power systems. 
Moreover, robust DNR with RG resizing can yield less conservative solutions, as the resizing operations can be proactively controlled. 
For instance, RG resizing can be implemented by installing mobile energy storage to store excessive energy when the output of RGs exceeds grid capacity or demand \cite{denholm2019timescales, rogers2010examples, chen2022robust}.
Previous research showed that the timescale of resizing energy storage is around 8 to 19 hours, which is consistent with that of network reconfiguration \cite{denholm2019timescales}. 
In Spain, the RG resizing is considered together with real-time power curtailments \cite{rogers2010examples}. 
In \cite{chen2022robust}, the resizing of RGs (called preparatory power curtailment) was performed to minimize the real-time curtailment and enhance grid stability in the economic dispatch.

In this paper, we address robust DNR with RG resizing, where the uncertainty set of RGs is influenced by the here-and-now decision, a property known as decision-dependent uncertainty (DDU) \cite{ben2004adjustable}.
Therefore, the robust DNR problem considering RG resizing is modeled as a two-stage RO with DDU.
Different from decision-independent uncertainty (DIU), DDU presents great challenges due to the mutual influence between uncertainties and decisions \cite{nohadani2018optimization, zhang2022two}.
Classical two-stage RO algorithms are often inapplicable to DDU, as the previously selected scenarios may fall outside the uncertainty set when the here-and-now strategy changes. 
DDU has gained considerable attention, with numerous algorithms to tackle the difficulties in the two-stage RO with DDU \cite{ nohadani2018optimization, zhang2022two, feng2021multistage, yang2024robust, xie2024robust, zeng2022two}. 
In general, these studies can be divided into two categories. 
In the first category, the DDU-based uncertainty set is reformulated as a DIU-based counterpart by establishing a bijection between the associated groups of uncertainty variables \cite{feng2021multistage, yang2024robust, xie2024robust}.
For example, reference \cite{xie2024robust} solved the two-stage RO with DDU by constructing new uncertainty variables in a DIU set and returning to the master problem the constraints characterizing the relationship between RG resizing variables and the worst-case scenarios. 
The basis of constraints, which depends on the obtained worst-case scenario in the uncertainty set, was returned to the master problem in the basic C\&CG framework in \cite{chen2022robust}.
An alternative approach leverages the fact that the feasible region of the dual variable \( \lambda \)  is independent of the first-stage decision \( \xi \), and returns only the worst \( \lambda^{\ast} \) to the master problem \cite{zeng2022two}.
Specifically, by applying the double duality operation, the uncertainty variable \(w\) is reformulated from the outer maximization problem into the inner minimization problem.
Reference \cite{zhang2022two} studied a two-stage linear RO with polyhedral DDU-based uncertainty sets and proposed a modified Benders decomposition algorithm to solve it.

However, previous literature only focused on the DDU-based two-stage RO with linear constraints for tractability and guarantee of convergence.
In this paper, we consider a nonlinear power flow-constrained robust DNR problem with RG resizing, which acts as a here-and-now decision that changes the structure of the uncertainty set and introduces decision dependence.
The incorporation of nonlinear constraints brings difficulties in computations to existing methods, for instance, it prevents the modified Benders decomposition algorithm \cite{zhang2022two} from converging in a finite number of steps.
To efficiently solve the problem considered, a customized mapping-based C\&CG is developed.

Our main contributions can be summarized as follows:
\begin{itemize}
    \item  The RG resizing is considered in the robust DNR problem, co-optimized together with the network topology in the first stage, which is modeled as a two-stage robust mixed-integer nonlinear RO with DDU.
    Distinct from the traditional robust DNR problem \cite{lee2014robust}, the variation in the RG output is bounded by here-and-now decision--RG resizing, introducing DDU to the uncertainty set. 
    \item The DDU in the robust DNR problem brings great difficulty in solving it with traditional algorithms.
    We develop a customized mapping-based C\&CG algorithm, to solve the problem. 
    The proposed algorithm can converge to the optimal solution in finite iterations when the uncertainty set is a polyhedron. 
    Case studies validate the optimality of the solution and the finite-step convergence.
    Moreover, the results show that our algorithm can greatly reduce computational complexity compared to the modified Benders dual decomposition \cite{zhang2022two}.
    \item  The sensitivity of the optimization result with respect to the uncertainty set parameters in the two-stage RO with DDU is theoretically analyzed for the first time, which is performed in two steps.
    First, the sensitivity of the second-stage optimal value for a fixed first-stage decision is analyzed by the corresponding dual variables. 
    Second, the impact of the uncertainty set parameters on the first-stage optimal solution is analyzed through the structure of the robust DNR.
\end{itemize}

The rest of this paper is organized as follows. 
Section \ref{section2} introduces the mathematical formulation of the robust DNR problem. The mapping-based C\&CG algorithm is developed in Section \ref{section3} to solve this two-stage RO with DDU. The sensitivity with respect to the uncertainty set parameters is analyzed in Section \ref{section4}. Case studies are reported in Section \ref{case_study}, with conclusions in Section \ref{section6}.

\section{Mathematical Formulation}\label{section2}

We consider the robust DNR problem with bus set $\mathcal{N}$, branch set $\mathcal{E}$ and dispatch period set $\mathcal{T}$. 
The distribution network is equipped with \( N_g \) thermal generators, \( N_w \) RGs and \( N_e  \) battery energy storages (BESs). 
The distribution network is modeled as a directed graph \( \mathcal{G} ( \mathcal{N}, \mathcal{E} ) \), where each branch in $\mathcal{E}$ is arbitrarily assigned a reference direction from bus \(i\) to bus \(j\), denoted as $(i,j) \in\mathcal{E}$. 
A binary variable \(\alpha_{ij}\) represents the switch status of the branch \( (i,j) \), in which \( \alpha_{ij} = 0  \) if the branch \( (i,j) \) is open, otherwise \( \alpha_{ij} =1 \). 
With the setting above, we formulate the DNR problem
with RG resizing as follows.

\subsection{Distribution Network Constraints} 
For simplicity, we omit the notation \(t\) at the subscript of the second-stage variables.

\subsubsection{Power balance}
The active and reactive power balance constraints at each bus \( j \in \mathcal{N} \) and each time \( t \in \mathcal{T} \) are:
\begin{subequations}\label{eq:power}
\begin{eqnarray}
\sum_{(i,j)\in \mathcal{E}} \left ( p_{ij,t}-r_{ij}l_{ij,t}  \right ) + p_{j,t} &=&  \sum_{(j,k)\in \mathcal{E}}  p_{jk,t},\nonumber \\
&&~~\forall j \in \mathcal{N},~t \in \mathcal{T}  \label{eq:pj}\\
\sum_{(i,j)\in \mathcal{E}} \left ( q_{ij,t}-x_{ij}l_{ij,t}  \right ) + q_{j,t} &=&  \sum_{(j,k) \in \mathcal{E}}  q_{jk,t},\nonumber\\
&&~~ \forall j \in \mathcal{N},~t \in \mathcal{T} \label{eq:qj} 
\end{eqnarray}
\end{subequations}
in which \( l_{ij,t} \) denotes the square of the current of branch \( (i,j) \),  \( p_{j,t} = w_{j,t} + p_{j,t}^{bes} +p_{j,t}^{G} -  p_{j,t}^{L}   \), \( q_{j,t} = q_{j,t}^{G} - q_{j,t}^{L}  \) are active and reactive power injections at bus \(j\), and constant \( r_{ij}, x_{ij} \) denote the resistance and reactance, respectively.
The load power \( (p_{j,t}^{L}, q_{j,t}^{L}) \) is given, while the RG output \( w_{j,t} \), the BES power \( p_{j,t}^{bes} \)  and the thermal generation \( (p_{j,t}^{G}, q_{j,t}^{G}) \) are variables to be introduced shortly.

\subsubsection{Branch power flow}
The branch power flow at each branch \( (i,j)  \) after a convex relaxation \cite{peng2014feeder} can be described by:  
\begin{eqnarray}\label{eq:branch_relax}
 p_{ij,t}^2 + q_{ij,t}^2  \le  l_{ij,t}   v_{i,t}, \quad \forall (i,j) \in \mathcal{E}, t \in \mathcal{T}
\end{eqnarray}
in which \( v_{i,t} \) denotes the square of voltage of bus \( i\).

\subsubsection{Voltage difference between buses }
By using big-M method, the voltage difference between two buses \(i\) and \(j\) of each branch \((i,j)  \) at any time period \(t\) is:
\begin{eqnarray}\label{eq:voltage}
   v_{i,t} - v_{j,t} - M\left ( 1 - \alpha _{ij} \right )    & \le &   \nonumber  \\   
  2\left ( r_{ij} p_{ij,t} + x_{ij}q_{ij,t} \right ) &-& \left ( r_{ij}^2 + x_{ij}^2 \right ) l_{ij,t}  \\ 
 & \le &   v_{i,t} - v_{j,t}  + M\left ( 1 - \alpha _{ij} \right )    \nonumber 
\end{eqnarray}
in which \(M\) is a big positive number. 

\subsubsection{Safety limits} The bus voltages, branch currents, active and reactive power flows are limited as follows: 
\begin{subequations}\label{eq:safe_limits}
\begin{eqnarray}
  v_i^{\text{min}}   \le   v_{i,t} \le    v_i^{\text{max}}     \qquad  \forall i \in \mathcal{N} , t \in \mathcal{T},  &&\label{eq:voltage_limit}\\
0 \le l_{ij,t}  \le  \alpha _{ij} l_{ij} ^{\text{max}}   \quad  \forall (i,j) \in \mathcal{E} , t \in \mathcal{T}, && \label{eq:current_limit}  \\
- \alpha _{ij} p_{ij}^{\text{min}} \le p_{ij,t} \le \alpha _{ij} p_{ij}^{\text{max}} \quad  \forall (i,j) \in \mathcal{E}  ,  t \in \mathcal{T},   && \label{eq:active_limit}  \\
- \alpha _{ij} q_{ij}^{\text{min}} \le q_{ij,t} \le 
\alpha _{ij} q_{ij}^{\text{max}} \quad  \forall (i,j) \in \mathcal{E} , t \in \mathcal{T}  .  && \label{eq:reactive_limit}  
\end{eqnarray}
\end{subequations}

\subsection{Power Generation Limitation} 
The power imported at  bus \( i \in \mathcal{N}_g \) equipped with a thermal generator or \( i \in \mathcal{N}_e \) with a BES is restricted as \eqref{eq:injection_limits}:
\begin{subequations}\label{eq:injection_limits}
\begin{eqnarray}
\underline{p}_i^{G} \le  p_{i,t}^{G} \le \bar{p}_i^{G} \qquad  \forall i \in \mathcal{N}_g    , t \in \mathcal{T},  &&\label{eq:pG_limit}\\
\underline{q}_i^{G} \le q_{i,t}^{G} \le \bar{q}_i^{G}  \qquad  \forall i \in \mathcal{N}_g, t \in \mathcal{T},  && \label{eq:qG_limit}  \\
\underline{p}_i^{bes}   \le p_{i,t}^{bes} \le \bar{p}_i^{bes} \quad  \forall i \in \mathcal{N}_e , t \in \mathcal{T}.  && \label{eq:soc_limit} 
\end{eqnarray}
\end{subequations}

The state of charge (SoC) of each BES  should not exceed upper and lower bounds at any time:
\begin{eqnarray}\label{eq:soc}
S_{i}^{\text{min}}  \le  S_{i}(0)  - \sum_{t'  = 1}^{t}   p_{i,t'}^{bes}\triangle t   \le  S_{i}^{\text{max}}, ~\forall i \in \mathcal{N}_e, t \in \mathcal{T}, 
\end{eqnarray}
in which \( S_{i}(0) \) is the initial SoC of BES \(i\). \( S_{i}^{\text{min}} \) and \(S_{i}^{\text{max}}\) are the lower and upper bounds of the SoC of BES \(i\). \( \triangle t \) is the time interval. 

\subsection{Network Topology Constraint} 
The distribution network must maintain a radial structure, which is equivalent to the following two conditions:
\begin{enumerate}
    \item There are \( N - N_s \) closed branches in the network, in which \( N_s\) is the number of substation buses.
    \item There is no connected power loop in the network.
\end{enumerate}

Without loss of generality, we assume there is one substation bus in the distribution network (\( N_s  = 1 \)).
The following linear constraints \eqref{equation-9} on the switching status \( \boldsymbol{x}=(\alpha_{ij},\forall (i,j) \in \mathcal{E}) \) can be derived.
\begin{eqnarray}\label{equation-9}
\sum_{(i,j) \in \mathcal{E}}  \alpha_{ij} &=& N - 1, \\
\sum_{(i,j) \in \mathcal{L}_k} \alpha_{ij}   &\le & M_k - 1 , \quad  \forall ~  k \in \mathcal{K},  \nonumber
\end{eqnarray}
in which \(\mathcal{L}_k\) is the \(k^{\textnormal{th}}\) power loop in the network and \( M_k \) is the number of branches in this loop. 
The set \(\mathcal{K}\) is a prespecified set of power loops.
For a distribution network with one substation bus, power loops are equivalent to circuits in the network. 
There exist various algorithms to identify these loops, such as the depth-first search algorithm and Johnson's algorithm \cite{mateti1976algorithms}.
Let \(  \mathcal{X}  := \left \{  \boldsymbol{x} | \boldsymbol{x} \text{ satisfies } \eqref{equation-9} \right \}  \) denote the set of all feasible network topologies.

\subsection{Resizing the Renewable Generators} 

This paper considers the resizing of RGs in the first stage to reduce frequent real-time curtailment and stabilize the grid in response to higher renewable energy penetration and increasing power fluctuations.
The RG output limits are resized through the allocation of energy storage systems, which store excessive renewable energy and limit the RG outputs.
The resizing variable is treated as an upper bound of the actual RG output, denoted as \( \boldsymbol{\xi} = (\xi_{i,t},  \forall i,  \forall t) \).
The uncertainty sets without and with RG resizing are denoted as \( \hat{\mathcal{W}}   \) and \( \mathcal{W}(\boldsymbol{\xi})  \), respectively.
The set of natural RG outputs \( \hat{\mathcal{W}} \) is constrained as follows:
\begin{subequations}\label{uncertain_constraint}
\begin{eqnarray}
w_{i,t}^{\text{min}} \le \hat{w}_{i,t} \le w_{i,t}^{\text{max}}, ~ \forall i \in \mathcal{N}_w, t \in \mathcal{T},   \label{uncertain_constrainta} \\
\sum_{i \in \mathcal{N}_w} \frac{\left | \hat{w}_{i,t} - w_{i,t}^p   \right |}{\Delta w_{i,t}   }  \le \Gamma_t,  ~ \forall t \in \mathcal{T},  \label{uncertain_constraintb}\\
\sum_{t = 1}^{T} \frac{\left | \hat{w}_{i,t} - w_{i,t}^p   \right |}{\Delta w_{i,t}}  \le \Gamma_i,  ~\forall i \in \mathcal{N}_w \label{uncertain_constraintc}
\end{eqnarray}
\end{subequations}
in which \(\hat{w}_{i,t} \) is the output of RG \(i\) before resizing at time \(t\), \( w_{i,t}^p := \frac{1}{2}   ( w_{i,t}^{\text{min}} + w_{i,t}^{\text{max}} ) \), \(  \Delta w_{i,t} := \frac{1}{2}   ( w_{i,t}^{\text{max}} - w_{i,t}^{\text{min}}) \),  \( w_{i,t}^{\text{max}} \) and \( w_{i,t}^{\text{min}} \) are its upper and lower bounds, respectively.
Constraints \eqref{uncertain_constraintb} and \eqref{uncertain_constraintc} limit the deviation of the unresized RG outputs from the given forecast values \(w_{i,t}^p\) in temporal and spatial aspects by the positive constants \( \Gamma_t \) and \( \Gamma_i \), respectively.

Let \( \boldsymbol{w} = ( w_{i,t}, \forall i, t ) \) denote the output of RGs with resizing. 
The uncertainty set after RG resizing can be written as follows:
\begin{eqnarray}\label{uncertain_set}
\mathcal{W}(\boldsymbol{\xi})  =  \left \{  \boldsymbol{w} \mid w_{i,t} = \text{min}\{ \hat{w}_{i,t},  \xi_{i,t} \} ,\forall i,   t  , ~     \hat{\boldsymbol{w}} \in  \hat{\mathcal{W}}      \right \}  . 
\end{eqnarray}
We limit the resizing variable \( \boldsymbol{\xi} \) within the upper bounds of RG outputs \( \boldsymbol{w}^{ \max } := \{ w_{i,t}^{\max}, ~ \forall i \in \mathcal{N}_w, ~ \forall t \in \mathcal{T}  \} \) as below:
\begin{eqnarray}\label{curtailment}
 \boldsymbol{\xi} \le   \boldsymbol{w}^{ \max }   .
\end{eqnarray}

\subsection{Robust DNR Problem} 
Our objective is to find a radial network topology that minimizes the total cost of branch switching, RG resizing, and power injections from BESs and thermal generators. 
We consider the following cost:
\begin{subequations}
\begin{eqnarray}\label{obj}
&&\sum_{(i,j) \in \mathcal{E}    } c_{\alpha, ij} \alpha _{ij} -\sum_{t \in \mathcal{T}    }   \sum_{i \in \mathcal{N}_w  }  c_{\xi, i,t} \xi_{i,t} \label{object_a}
\\  &+& \sum_{t \in \mathcal{T}    }   \sum_{i \in \mathcal{N}_e    } b_{i}^{bes} p_{i,t}^{bes} + \sum_{t \in \mathcal{T}    }   \sum_{i \in \mathcal{N}_g    }
\left (  b_{i}^{p} p_{i,t}^{G} +   b_{i}^{q} q_{i,t}^{G} \right ) \label{object_b}
\end{eqnarray}
\end{subequations}
in which \( c_{\alpha, ij} \), \(  c_{\xi, i,t} \), \( b_{i}^{bes} \), \( b_{i}^{p} \), \(  b_{i}^{q} \) are constant factors for the cost of switching, RG resizing, BESs, active power generation, and reactive power generation, respectively.
Collect the variables related to power injections and flows as \( \boldsymbol{y} := \left \{ p_{i,t}^{bes},  p_{i,t}^{G},  q_{i,t}^{G}, v_{i,t},  l_{ij,t}, p_{ij,t}, q_{ij,t},  \forall i \in  \mathcal{N},  (i,j) \in \mathcal{E} ,  t \in \mathcal{T}   \right \} \).
We formulate the robust DNR (RDNR) problem as follows:
\begin{eqnarray}\label{P：DNR}
\textbf{RDNR}:  \min_{\boldsymbol{x} \in \mathcal{X},~ \boldsymbol{\xi} \le \boldsymbol{w}^{ \max }} \left \{  \boldsymbol{c}_{\alpha}^{\top} \boldsymbol{x}\! -\! \boldsymbol{c}_{\xi}^{\top} \boldsymbol{\xi} \!+ \!\max_{\boldsymbol{w} \in \mathcal{W}(\boldsymbol{\xi})} Q(\boldsymbol{x},\! \boldsymbol{w}) \right \} 
\end{eqnarray}
in which \(  Q(\boldsymbol{x}, \boldsymbol{w}) \) denotes the optimal value of
\begin{subequations}\label{eq:inner_minimization}
\begin{eqnarray}\label{Qfunction}
&& \min_{\boldsymbol{y}}  ~~ \boldsymbol{b}^{\top} \boldsymbol{y}      \label{Qfunctiona}   \\
&& \text{ s. t. } ~   A\boldsymbol{x} + B\boldsymbol{y} + \gamma \le G \boldsymbol{w} \label{Qfunctionb} \\ 
&&  \qquad ~~ \left \| C_{ij,t} \boldsymbol{y} \right \|_2 \le d_{ij,t}^{\top} \boldsymbol{y}, \forall (i,j) \in \mathcal{E}, ~t \in  \mathcal{T} \label{Qfunctionc}
\end{eqnarray} 
\end{subequations}
in which \( \boldsymbol{b}^{\top} \boldsymbol{y}  \) corresponds to \eqref{object_b}. 
The linear constraints  \eqref{eq:power}, \eqref{eq:voltage}-\eqref{eq:soc} are collected as \eqref{Qfunctionb}, while the relaxed branch power flow constraint \eqref{eq:branch_relax} is converted to a second-order cone \eqref{Qfunctionc} by \cite{farivar2013branch}.
The first-stage RG resizing variable \(\boldsymbol{\xi}\) affects the uncertainty set \( \mathcal{W}(\boldsymbol{\xi}) \) and further influences \(Q(\boldsymbol{x}, \boldsymbol{w}) \) through \( \boldsymbol{w} \). 
Furthermore, let \( \mathcal{Y}(\boldsymbol{x}, \boldsymbol{w})  \) denote the feasible region of \( \boldsymbol{y} \) for a given \( (\boldsymbol{x}, \boldsymbol{w}) \).
\begin{eqnarray}
\mathcal{Y}(\boldsymbol{x}, \boldsymbol{w}) := \{ \boldsymbol{y} \mid  \boldsymbol{y} \text{ satisfies } \eqref{Qfunctionb}, \eqref{Qfunctionc} \text{ under }(\boldsymbol{x}, \boldsymbol{w})  \}.
\end{eqnarray}

\section{Solution Algorithm}\label{section3}

The RDNR problem above is a two-stage mixed-integer robust optimization problem, in which the second stage is a second-order cone program (SOCP).
Let \( \mathcal{R} \) be the feasible region of the first-stage strategy \( \left (   \boldsymbol{x},  \boldsymbol{\xi}   \right )  \) defined as: 
\begin{eqnarray}
\mathcal{R}   &=&   \{  (\boldsymbol{x}, \boldsymbol{\xi} )    \mid  \boldsymbol{x} \in \mathcal{X} , \boldsymbol{\xi} \le \boldsymbol{w}^{ \max},  \nonumber \\
&&  \quad   \text{ and } \mathcal{Y}(\boldsymbol{x}, \boldsymbol{w}) \neq \varnothing, \forall \boldsymbol{w} \in \mathcal{W}(\boldsymbol{\xi})   \} .  \nonumber
\end{eqnarray}
Unlike traditional two-stage RO models \cite{lee2014robust}, the uncertainty set \( \mathcal{W}(\boldsymbol{\xi})\) \eqref{uncertain_set} depends on the first-stage strategy \( \boldsymbol{\xi} \), making it decision-dependent.
Traditional algorithms, e.g., Benders decomposition and C\&CG algorithm, directly return a worst-case uncertainty variable \( \boldsymbol{w}^{\ast} \) and generate the cutting plane through it. 
However, they cannot be directly applied to the two-stage RO with DDU because of the following challenges:
\begin{enumerate}
    \item The previous worst-case scenario may fall outside of the DDU set associated with a new first-stage decision, leading to infeasibility or overly restrictive solutions.
    \item When the uncertainty set is decision-dependent, the previous worst-case scenarios may no longer be vertices of the new set, compromising convergence guarantees.
\end{enumerate}

Next, we propose a customized algorithm called mapping-based C\&CG to tackle the RDNR problem \eqref{P：DNR} by generating a mapping from the first-stage decision \(  \boldsymbol{\xi} \) to a vertex of the decision-dependent uncertainty set.

\subsection{Mapping from Resizing Decision to Worst-case Scenario}\label{3-C}

From the above analysis, the key reason for the failure of the traditional C\&CG algorithm is that it returns a fixed scenario to the master problem in each iteration, while the uncertainty set itself changes with the first-stage decision.
This inspires us to find and return a mapping from the resizing variable \( \boldsymbol{\xi} \) to the worst-case scenario \( \boldsymbol{w}^{\ast} \) under a given topology \( \boldsymbol{x} \).

The uncertainty set with RG resizing \( \mathcal{W}(\boldsymbol{\xi}) \) is transformed to the following form: 
\begin{eqnarray}\label{dual_uncertainty_set}
\mathcal{W}(\boldsymbol{\xi}) &=&
\left\{\boldsymbol{w} ~ \middle|\
\begin{array}{l}
w_{i,t}^{\text{min}} \le w_{i,t} \le w_{i,t}^{\text{max}}, ~ \forall i \in \mathcal{N}_w, t \in \mathcal{T} , \\ 
\displaystyle \sum_{i \in \mathcal{N}_w} \frac{\left | w_{i,t} - w_{i,t}^p   \right |}{\Delta w_{i,t}   }  \le \Gamma_t,  ~ \forall t \in \mathcal{T}, \\[2ex]
\displaystyle \sum_{t = 1}^{T} \frac{\left | w_{i,t} - w_{i,t}^p   \right |}{{\Delta w_{i,t} }}  \le \Gamma_i,  ~\forall i \in \mathcal{N}_w, \\[2ex]
 w_{i,t} \le \xi_{i,t}, ~ \forall i \in \mathcal{N}_w, t \in \mathcal{T} 
\end{array}
\right\}\nonumber \\
&=& \left\{ \boldsymbol{w} \mid F \boldsymbol{w} \le f, \boldsymbol{w} \le  \boldsymbol{\xi}    \right\}  .
\end{eqnarray}
in which \( F \boldsymbol{w} \le f  \) represents the first three groups of constraints and \(  \boldsymbol{w} \le  \boldsymbol{\xi} \) represents the last constraint.
The new description of the uncertainty set \eqref{dual_uncertainty_set} imposes a group of upper bound constraints on the natural RG outputs \( \hat{\boldsymbol{w}} \), which represent the changes caused by RG resizing. 
\begin{remark}\label{remark11}
    The uncertainty set \eqref{dual_uncertainty_set} is equivalent to the original set \eqref{uncertain_set} if the resizing decision \(\boldsymbol{\xi}\) satisfies:
\begin{subequations}\label{eq:remark1}
\begin{eqnarray}
&&   \boldsymbol{w}^{\min} \le \boldsymbol{\xi}   \\
&&   \displaystyle \sum_{i \in \mathcal{N}_w} \frac{  w_{i,t}^p - \xi_{i,t} }{\Delta w_{i,t}  }  \le \Gamma_t,  ~ \forall t \in \mathcal{T}, \\
&&   \displaystyle \sum_{t = 1}^{T} \frac{ w_{i,t}^p - \xi_{i,t}  }{{\Delta w_{i,t} }}  \le \Gamma_i,  ~\forall i \in \mathcal{N}_w
\end{eqnarray}
\end{subequations}
i.e., the resizing variable \( \boldsymbol{\xi} \) must respect the lower bound of the uncertainty set \( \mathcal{W}(\boldsymbol{\xi}) \).
In fact, if \( \boldsymbol{\xi} \) were to fall below the minimum specified in \eqref{eq:remark1}, the transformed uncertainty set in \eqref{dual_uncertainty_set} collapses to an empty set, rendering the problem infeasible.
In practice, system operators schedule generation according to forecasts of load and renewable output \cite{chen2022robust}. 
Consequently, large downward adjustments of generation resources are rarely required, and the constraints \eqref{eq:remark1} for \( \boldsymbol{\xi} \) are realistic.
\end{remark}

Let \(\Xi\) denote the feasible set defined by \eqref{curtailment} and \eqref{eq:remark1}.
Given any first-stage decision \( (\boldsymbol{x}, \boldsymbol{\xi} ) \), we consider the maximization problem over uncertainty:
\begin{eqnarray}\label{SSOP}
&&\max_{\boldsymbol{w} \in \mathcal{W}(\boldsymbol{\xi}) } ~~ Q(\boldsymbol{x}, \boldsymbol{w})  
\end{eqnarray}

Few analytical methods can be directly applied to \eqref{SSOP} since its objective \( Q(\boldsymbol{x}, \boldsymbol{w})\) to be maximized is convex in \( \boldsymbol{w} \).
This implies that the maximizer \( \boldsymbol{w}^{\ast} \) must lie at an extreme point of the feasible set \( \mathcal{W}(\boldsymbol{\xi}) \).
Taking the dual of the minimization problem \eqref{eq:inner_minimization} that defines \( Q(\boldsymbol{x}, \boldsymbol{w})  \), the problem \eqref{SSOP} can be equivalently transformed into the following form:
\begin{eqnarray}\label{first-dual}
&& \max_{ \boldsymbol{w} \in \mathcal{W}(\boldsymbol{\xi}) ,\boldsymbol{\lambda} \ge 0 ,\boldsymbol{\mu} } ~~  \lambda_t^{\top }   \left ( A\boldsymbol{x}  + \gamma  \right )  - \lambda_t^{\top } G \boldsymbol{w}   \nonumber \\
&&  \quad  \text{ s. t. } ~~ \quad B^{\top }\lambda_t  = C^{\top }\boldsymbol{\mu} + D^{\top } \lambda_h   \\
&&  ~~  \qquad   \qquad \left \| \mu _{ij,t} \right \|_2 \le \lambda _{h,ij,t},  ~ \forall (i,j) \in \mathcal{E},  t \in  \mathcal{T}    \nonumber 
\end{eqnarray}
in which \(  \lambda_t, \lambda_h  \) are dual variables for constraints \eqref{Qfunctionb}, \eqref{Qfunctionc}, respectively, and \( \boldsymbol{\mu} \) is the dual variable for an auxiliary equality that substitutes the term $(C_{ij,t}\boldsymbol{y})$ with a single variable.
The coefficient matrices \( C \) and \(D\) are constructed by respectively stacking  \(  C_{ij,t} \) and \( d_{ij,t} \) row-wise.
The objective function of \eqref{first-dual} contains a bilinear term \( \lambda_t^{\top } G \boldsymbol{w} \).
Noticing that the uncertainty set \( \mathcal{W}(\boldsymbol{\xi}) \) is polyhedral under any feasible \(  \boldsymbol{\xi} \), its extreme points can be represented with binary variables, and the bilinear term \( \lambda_t^{\top } G \boldsymbol{w} \) can be linearized by the big-M method \cite{lee2014robust}.

Given an optimal dual solution \(   (\boldsymbol{\lambda}^{\ast},  \boldsymbol{\mu}^{\ast}) \), the worst-case scenario \(  \boldsymbol{w}^{\ast} \) is obtained by solving the linear program:
\begin{eqnarray}\label{max_w}
&&  \max_{\boldsymbol{w} } ~~ -(G^{\top } \lambda_t^{\ast}  )^{\top } \boldsymbol{w}  \nonumber \\
&&~\text{s. t.}  \quad F\boldsymbol{w} \le f :  \boldsymbol{\pi} \\
&&\qquad \quad   \boldsymbol{w} \le \boldsymbol{\xi} : \boldsymbol{\theta} \nonumber
\end{eqnarray}
Considering the Karush-Kuhn-Tucker (KKT) condition of \eqref{max_w}, we can get the set of the worst-case scenarios in \eqref{kkt-condition}.
\begin{eqnarray}\label{kkt-condition}
\mathcal{K}  (\boldsymbol{\xi}, \lambda_t^{\ast}) = \left\{  (\boldsymbol{w}, \boldsymbol{\pi}, \boldsymbol{\theta})   \middle|
\begin{array}{l}
G^\top \lambda_t^{\ast} + F^\top \boldsymbol{\pi} + \boldsymbol{\theta} = 0 \\
F\boldsymbol{w}  \leq f  \quad \boldsymbol{w}  \leq \boldsymbol{\xi} \\
0 \leq \boldsymbol{\pi} \quad  \quad 0 \leq \boldsymbol{\theta} \\
\boldsymbol{\pi} \circ  (F\boldsymbol{w}  - f) = 0 \\
\boldsymbol{\theta} \circ (\boldsymbol{w}  - \boldsymbol{\xi}) = 0
\end{array}
\right \}
\end{eqnarray}
in which the symbol \(  x  \circ  y  \) denotes the Hadamard (element-wise) product of vectors \(x\) and \(y\).
In \eqref{kkt-condition}, lines 2-3 represent primal and dual feasibility.
Lines 4-5 are the complementary slackness, in which a non-zero \( \pi_i \) (or \( \theta_i \)) implies an active constraint \( F_i \boldsymbol{w} = f_i  \) (or \( \boldsymbol{w}_i = \boldsymbol{\xi}_i\)).

\begin{remark}
    The set \( \mathcal{K}  (\boldsymbol{\xi}, \lambda_t^{\ast}) \) may not be a singleton with respect to \((\boldsymbol{\pi},\boldsymbol{\theta})\) even when the optimal \( \boldsymbol{w}^{\ast} \) of \eqref{max_w} is unique.
    Such non-uniqueness arises when the active constraints corresponding to the optimal \( \boldsymbol{w}^{\ast} \) are redundant. 
    Nevertheless, the set $\mathcal{K}  (\boldsymbol{\xi}, \lambda_t^{\ast})$ in \eqref{kkt-condition} still characterizes the unique $\boldsymbol{w}^{\ast}$ and thus induces $\mathcal{K}  (\boldsymbol{\xi}, \lambda_t^{\ast})$ as a mapping: $\boldsymbol{\xi}\mapsto  \boldsymbol{w}^{\ast}$.
    This mapping explicitly describes how the worst-case realization would adapt to the change of the uncertainty set caused by the first-stage RG resizing decision.
\end{remark}

To illustrate how the first-stage decision \( \boldsymbol{\xi} \) influences \( \boldsymbol{w}^{\ast} \) through the mapping in \eqref{kkt-condition}, we present a toy example with two RGs and a single time period. 
Consider the following decision-dependent uncertainty set:
\begin{eqnarray}
\mathcal{W}(\boldsymbol{\xi}) &=& \left\{ (w_1, w_2)^\top \;\middle|\;
\begin{array}{l}
w_j \le \xi_j , \quad \forall j = 1, 2 \\
0.25 \leq  w_j \leq 0.75, \quad \forall j = 1, 2 \\
\displaystyle \frac{\left | w_1 - 0.5 \right |  }{0.25} + \frac{\left | w_2 - 0.5 \right | }{0.25} \leq 1
\end{array}
\right\} \nonumber
\end{eqnarray}


\begin{figure}
\centering
\subfigure[]{
\includegraphics[width=0.51\columnwidth]{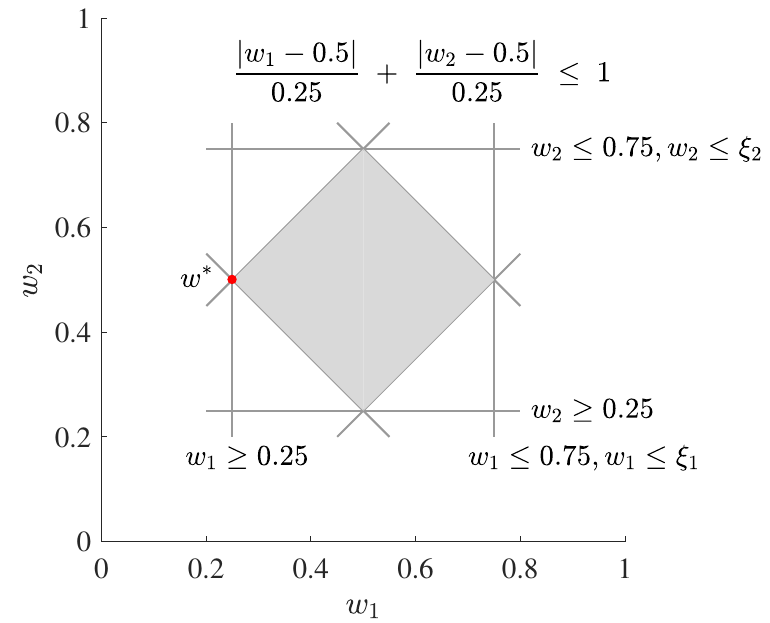}
}
\hfill
\subfigure[]{
\includegraphics[width=0.42\columnwidth]{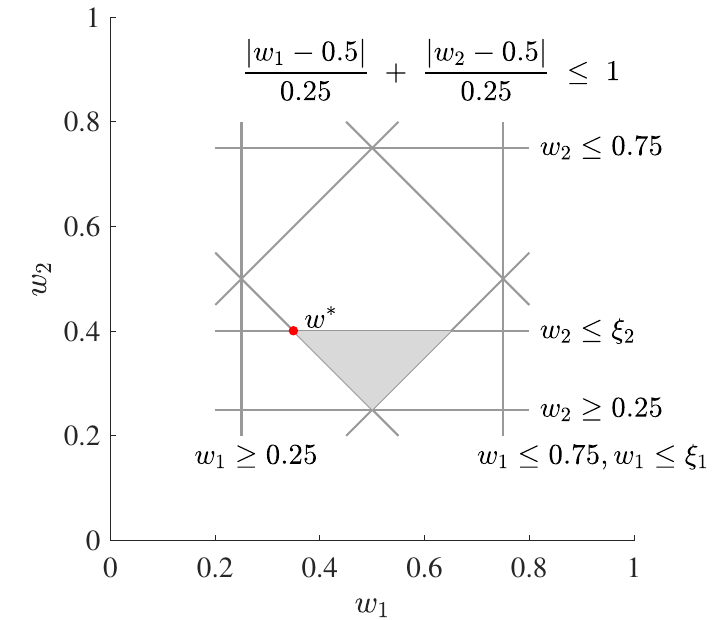}
}
\caption{Illustration of the KKT-based mapping \eqref{kkt-condition}.}	
 \label{fig:mapping}
\end{figure}

Suppose the current resizing decisions are \( \xi_1 = \xi_2 = 0.75 \).
The uncertainty set \(  \mathcal{W}(\boldsymbol{\xi})   \) under this \( (\xi_1, \xi_2) \) is shown as the gray region in Fig. \ref{fig:mapping}(a).
The factors  \( G^\top \lambda_t^{\ast} \) are computed as \( (0.79, 0.63) \), so that the worst-case \( \boldsymbol{w}^{\ast}\) in Fig. \ref{fig:mapping}(a) is \( (0.25, 0.5) \).
When the resizing decision \( \xi_2 \) is adjusted to \(0.4\), the uncertainty set \(  \mathcal{W}(\boldsymbol{\xi})   \) changes to the one in Fig. \ref{fig:mapping}(b) and the worst-case $\boldsymbol{w}^{\ast}$  becomes \( ( 0.35, 0.4) \).
The worst-case $\boldsymbol{w}^{\ast}$ is a vertex of \(  \mathcal{W}(\boldsymbol{\xi})   \) that can change with \( \boldsymbol{\xi} \).

\subsection{Mapping-based C\&CG Algorithm}

We propose a mapping-based C\&CG algorithm,
which follows the master-subproblem framework and generates the variables and constraints to strengthen the master problem in each iteration.
The key innovation of the mapping-based C\&CG is that it returns the mapping \( \mathcal{K}  (\boldsymbol{\xi}, \lambda_t^{\ast}) \) from the first-stage decision \(\boldsymbol{\xi}\) to the worst-case scenario \( \boldsymbol{w}^{\ast} \) rather than just returning a fixed \( \boldsymbol{w}^{\ast} \) in the traditional C\&CG algorithm.
The mapping \( \mathcal{K}  (\boldsymbol{\xi}, \lambda_t^{\ast}) \) can dynamically recognize new extreme points as \(\boldsymbol{\xi}\) changes.
Therefore, the mapping-based C\&CG method can conquer the challenges raised by DDU and converge to an optimal solution in a finite number of iterations.

It is noted that the RDNR problem does not naturally satisfy the relatively complete recourse property \cite{rockafellar1976stochastic}.
Consequently,
\( \max_{\boldsymbol{w} \in \mathcal{W}(\boldsymbol{\xi}) }  Q(\boldsymbol{x}, \boldsymbol{w})\) may not always be feasible for all first-stage decisions \( (\boldsymbol{x}, \boldsymbol{\xi} ) \), which results in unbounded maximum value and optimal $\boldsymbol{\lambda}$ of \eqref{first-dual} and thus the failure of commercial solvers to provide a valid solution.
To overcome this limitation, we analyze the robust feasibility condition as follows.

\subsubsection{Robust feasibility}
The robust feasibility of a first-stage decision \( \left (   \boldsymbol{x},  \boldsymbol{\xi}   \right )  \) can be examined by solving the following feasibility checking problem:
\begin{eqnarray}\label{feasible-checking}
&&\mathcal{F}\left (   \boldsymbol{x},  \boldsymbol{\xi}   \right )   := \max_{\boldsymbol{w} \in \mathcal{W}(\boldsymbol{\xi}) } \min_{\boldsymbol{\eta} \ge 0, \boldsymbol{y} }  \mathbf{1}  ^{\top} \boldsymbol{\eta}     \nonumber  \\
&& \qquad \qquad \quad \text{s. t.}~  A\boldsymbol{x} + B\boldsymbol{y} + \gamma \le G \boldsymbol{w} + \boldsymbol{\eta}   \\
     &&  \qquad \qquad   \qquad ~~ \left \| C_{ij,t} \boldsymbol{y} \right \|_2 \le d_{ij,t}^{\top} \boldsymbol{y},~\forall (i,j) \in \mathcal{E},  t \in  \mathcal{T}  \nonumber
\end{eqnarray}
in which the nonnegative slack variable \( \boldsymbol{\eta} \) relaxes the linear constraint.
Problem \eqref{feasible-checking} is always feasible due to existence of a trivial solution \(   \boldsymbol{y} =  \boldsymbol{0}    \), \(   \boldsymbol{\eta}  =  \max \{ \boldsymbol{0},  A\boldsymbol{x}  + \gamma -  G \boldsymbol{w} \} \).
If \( \mathcal{F}\left (   \boldsymbol{x},  \boldsymbol{\xi}   \right ) = 0 \), then \( \left (   \boldsymbol{x},  \boldsymbol{\xi}   \right ) \in \mathcal{R} \), i.e., it is robustly feasible.

Previous studies commonly add a feasibility cut based on the inner dual of \eqref{feasible-checking} to the master problem whenever \(\left ( \boldsymbol{x}, \boldsymbol{\xi} \right ) \) is not robustly feasible \cite{chen2022robust, zhang2022two, zeng2022two}.
However, adding such a feasibility cut is computationally inefficient for our problem due to the nonlinearity of the second stage.
Instead, we consider the following relaxed active power balance equation:
\begin{eqnarray}\label{relaxed}
\sum_{(i,j)\in \mathcal{E}} \left ( p_{ij,t}-r_{ij}l_{ij,t}  \right ) + p_{j,t} =  \sum_{(j,k)\in \mathcal{E}}  p_{jk,t}  +  s_{j,t}   
\end{eqnarray}
where the slack variable \( s_{j,t} \ge 0  \) compensates for any insufficient resizing \( \boldsymbol{\xi} \) to guarantee feasibility of the second-stage problem.
By substituting \eqref{eq:pj} with \eqref{relaxed}, we obtain a relaxed second-stage problem with relatively complete recourse:
\begin{eqnarray}\label{relaxed-2}
&& \max_{\boldsymbol{w} \in \mathcal{W}(\boldsymbol{\xi})} \min_{\boldsymbol{s} \ge 0,\boldsymbol{y}}  ~~ \boldsymbol{b}^{\top} \boldsymbol{y} +m_s \mathbf{1}^{\top}   \boldsymbol{s}   \nonumber  \\
&& ~ \text{ s. t. } ~   A\boldsymbol{x} + B\boldsymbol{y} + \gamma \le G \boldsymbol{w} + E\boldsymbol{s}  \\ 
&&  \qquad ~~ \left \| C_{ij,t} \boldsymbol{y} \right \|_2 \le d_{ij,t}^{\top} \boldsymbol{y}, ~\forall (i,j) \in \mathcal{E},  t \in  \mathcal{T}  \nonumber
\end{eqnarray} 
in which \( m_s \mathbf{1}^{\top} \boldsymbol{s}  \) is a penalty for violating the active power balance, with $m_s$ usually being a large positive factor.

\begin{remark}
For a robustly feasible \(  (\boldsymbol{x}, \boldsymbol{\xi} )    \), the relaxed second-stage model \eqref{relaxed-2} yields the same optimal solution and value as the original second-stage problem. 
When \(  (\boldsymbol{x}, \boldsymbol{\xi} )    \) is not robustly feasible, \eqref{relaxed-2} is still solvable and provides an approximate solution for the original formulation.
Therefore, we use \eqref{relaxed-2} as the subproblem to return the KKT-based mapping to the master problem.
\end{remark}

We next present the master problem and subproblem.

\subsubsection{Master problem (\textbf{MP})}
The \( k^{\textnormal{th}} \) iteration \textbf{MP} is
\begin{eqnarray}\label{master}
\mathbf{MP}:&&\min_{\boldsymbol{x},\boldsymbol{\xi} ,  \boldsymbol{\theta} \ge 0 , \boldsymbol{y}^i ,\boldsymbol{w}^i } \quad \boldsymbol{c}_{\alpha}^{\top} \boldsymbol{x} - \boldsymbol{c}_{\xi}^{\top} \boldsymbol{\xi} + L     \nonumber  \\
&&  \qquad~ \text{s. t.} \qquad  \boldsymbol{x}\in \mathcal{X} ,\boldsymbol{\xi} \in \Xi  \nonumber \\
&& \qquad \qquad \qquad L \ge  \boldsymbol{b}^{\top} \boldsymbol{y}^i, ~  \forall i \in \left [ k-1 \right ]      \\
&& \qquad \qquad \qquad \boldsymbol{y}^i \in \mathcal{Y}(\boldsymbol{x}, \boldsymbol{w}^i), ~ \forall i \in \left [ k -1\right ]  \nonumber    \\
&& \qquad \qquad \qquad    \boldsymbol{w}^i \in \mathcal{K}  (\boldsymbol{\xi}, \lambda_{t}^{i \ast}), ~ \forall i \in \left [ k -1 \right ]  .  \nonumber 
\end{eqnarray}
in which \(i\) is an iteration index, and \( \left [ k \right ] := \{ i: 1 \le i \le k, ~i \in \mathbb{Z}  \}  \).
Decision variables
\(  (\boldsymbol{y}^i,  \boldsymbol{w}^i) \) and their constraints are generated by the following subproblem \eqref{subproblem} in the \( i^{\textnormal{th}} \) iteration, with \( \lambda_{t}^{i \ast} \) denoting its optimal solution.

\subsubsection{Subproblem (\textbf{SP})}
Given solution \(  ( \boldsymbol{x}^k, \boldsymbol{\xi}^k ) \) of \textbf{MP} in the \(k^{\textnormal{th}}\) iteration, we have the following \textbf{SP} by the dual of \eqref{relaxed-2}.
\begin{eqnarray}\label{subproblem}
\mathbf{SP}:&& \max_{ \boldsymbol{w} \in \mathcal{W}(\boldsymbol{\xi}^k) ,\boldsymbol{\lambda} \ge 0 ,\boldsymbol{\mu} } ~~  \lambda_t^{\top }   \left ( A\boldsymbol{x}^k  + \gamma  \right )  - \lambda_t^{\top } G \boldsymbol{w}   \nonumber \\
&&  \quad  \text{ s. t. } ~~ \quad B^{\top }\lambda_t  = C^{\top }\boldsymbol{\mu} + D^{\top } \lambda_h   \\
&&  ~~  \qquad   \qquad  E^{\top} \lambda_t  \le m_s \mathbf{1}  \nonumber  \\
&&  ~~  \qquad   \qquad \left \| \mu _{ij,t} \right \|_2 \le \lambda _{h,ij,t}, ~ \forall (i,j) \in \mathcal{E}, t \in  \mathcal{T}     \nonumber 
\end{eqnarray}

\textbf{SP} \eqref{subproblem} can always attain a bounded optimal solution \( \lambda_t^{k \ast}   \) by virtue of the strong duality and feasibility of the primal problem \eqref{relaxed-2}.
Therefore, a mapping \( \mathcal{K}  (\boldsymbol{\xi}^k, \lambda_{t}^{k \ast})  \) can be generated and returned to the \textbf{MP} for the given \(  ( \boldsymbol{x}^k, \boldsymbol{\xi}^k )   \).

\subsubsection{Algorithm}
Based on \textbf{MP} and \textbf{SP}, we present a mapping-based C\&CG algorithm, Algorithm \ref{alg:DualCCG}, to solve RDNR \eqref{P：DNR}.
The proposed algorithm differs from the traditional C\&CG algorithm as the value of dual optimal \(  \lambda_{t}^{\ast}  \) is returned together with a mapping \(  \mathcal{K}  (\boldsymbol{\xi}, \lambda_{t}^{\ast})  \) from \( \boldsymbol{\xi}   \) to an extreme point of the uncertainty set $\mathcal{W}(\boldsymbol{\xi})$.

The \textbf{MP} \eqref{master} serves as a relaxation of, and provides a lower bound to, RDNR \eqref{P：DNR}.
By iteratively enumerating the mapping to the extreme point of the uncertainty set, the optimal value of \textbf{MP}  converges to that of the RDNR problem.
Noting that \(  Q(\boldsymbol{x}, \boldsymbol{w})  \) is convex in $\boldsymbol{w}$ and that its maximizer $\boldsymbol{w}^{\ast}$ must be an extreme point, the \textbf{MP} with all extreme points attached to its constraints becomes equivalent to RDNR \eqref{P：DNR}.
Therefore, Algorithm \ref{alg:DualCCG} can converge in a finite number of steps when the uncertainty set is a polyhedron or a finite set. The traditional C\&CG algorithm may fail to guarantee finite-step convergence, as conflicting uncertainty scenarios \( \boldsymbol{w}^{\ast } \) may be added to the master problem to make it infeasible.

\begin{algorithm}
\caption{Mapping-based C\&CG Algorithm}
\label{alg:DualCCG}
\begin{algorithmic}[1]
\renewcommand{\algorithmicrequire}{\textbf{Initialization:}}
\renewcommand{\algorithmicensure}{\textbf{Output:}}
\REQUIRE{Sufficiently large constant \(M>0\). Index \(k = 0\). Let \( {LB}_k = -M\), \({UB}_k = M\). Tolerance \(\varepsilon \).}
\STATE{ Solve \textbf{MP} \eqref{master} to obtain the optimal \((\boldsymbol{x}^k\), \(\boldsymbol{\xi}^k)\) and \(L_k\).}
 \STATE{Update \({LB}_k = \max \left \{  {LB}_{k-1},  \boldsymbol{c}_{\alpha}^{\top} \boldsymbol{x}^k  - \boldsymbol{c}_{\xi}^{\top} \boldsymbol{\xi}^k + L_k \right \}  \).}
 \STATE{Given \((\boldsymbol{x}^k,\boldsymbol{\xi}^k)\), solve \textbf{SP} \eqref{subproblem} to get the optimal \( \lambda_{t}^{k \ast} \) and the maximum objective value \( S_k \).}
 \STATE{Update \( {UB}_k = \min  \left \{  {UB}_{k-1} , c_{\alpha}^{\top} \boldsymbol{x}^k  - c_{\xi}^{\top} \boldsymbol{\xi}^k + S_k
 \right \}  \).}
\IF{  \({UB}_k - {LB}_k  \ge  \varepsilon \) }
    \STATE{ Create variables \(   (\boldsymbol{y}^k,   \boldsymbol{w}^k) \) and add the following
constraints to \textbf{MP} \eqref{master}: \\
\(\quad  L \ge  b^{\top} \boldsymbol{y}^k \) \\
  \( \quad  A\boldsymbol{x} + B\boldsymbol{y}^k + \gamma \le G \boldsymbol{w}^k, \)  \\
 \( \quad  \left \| C_{ij,t} \boldsymbol{y}^k \right \|_2 \le d_{ij,t}^{\top} \boldsymbol{y}^k,  ~ \forall (i,j) \in \mathcal{E},  \forall t \in  \mathcal{T}  \), \\
 \( \quad  \boldsymbol{w}^k \in \mathcal{K}  (\boldsymbol{\xi}, \lambda_{t}^{k^{\ast}})   \).}
    \STATE \( k = k + 1 \) and go back to Line 1.
\ENDIF
\STATE Terminate and output \( (\boldsymbol{x}^k, \boldsymbol{\xi}^k) \) as the optimal solution.
\end{algorithmic}
\end{algorithm}

\begin{remark}
    Without the decision-dependent constraint \(
\boldsymbol{w} \le \boldsymbol{\xi}  \), the KKT-based mapping \eqref{kkt-condition} will degenerate into
\begin{eqnarray}\label{degenerate}
\mathcal{K}  (\lambda_t^{\ast}) &=& \left\{
(\boldsymbol{w}, \boldsymbol{\pi})  \middle|
\begin{array}{l}
G^\top \lambda_t^{\ast} + F^\top \boldsymbol{\pi} = 0 \\
F\boldsymbol{w}  \leq f, \quad \boldsymbol{\pi} \ge 0 \\
\boldsymbol{\pi} \circ  (F\boldsymbol{w}  - f) = 0 
\end{array}
\right \}
\end{eqnarray}
which contains a single worst-case scenario \(  \boldsymbol{w}^{\ast}\).
In this case, the mapping-based C\&CG algorithm degenerates into the traditional C\&CG algorithm for solving an RO with DIU.
\end{remark}

\subsection{Comparison with Existing Algorithms} \label{discussion}
We next discuss the difference and advantages of the proposed mapping-based C\&CG algorithm, compared to existing algorithms for solving two-stage RO with DDU, including adaptive C\&CG \cite{chen2022robust}, nested C\&CG \cite{zhao2012exact}, and modified Benders dual decomposition \cite{zhang2022two}.

\subsubsection{Adaptive C\&CG \cite{chen2022robust}}
In this algorithm, after obtaining a worst-case scenario \( \boldsymbol{w}^{\ast} \) in the subproblem, it is necessary to identify the active constraints from the uncertainty set \( \mathcal{W}(\boldsymbol{\xi}) \) and return them to the master problem.
Some additional nonconvex constraints are appended to fix the potential degeneration issue in the process of active constraint identification, which makes this method complex and limits its applicability.

\subsubsection{Nested C\&CG \cite{zhao2012exact}}
The decision-dependent constraint \( w_{i,t} = \text{min}\{ \hat{w}_{i,t},  \xi_{i,t} \} \) can be linearized by introducing binary variables \(  z_{i,t} \in \{ 0, 1 \} \):
\begin{eqnarray}\label{nested_CCG}
&&w_{i,t} =  z_{i,t} \hat{w}_{i,t} + (1 - z_{i,t} )\xi_{i,t} , \nonumber \\
&&w_{i,t} \le \hat{w}_{i,t}, ~  w_{i,t} \le \xi_{i,t}, ~ \forall i, t  .
\end{eqnarray}
The RDNR problem \eqref{P：DNR} is turned into the following form.
\begin{eqnarray}
&&  \min_{\boldsymbol{x} \in \mathcal{X},~ \boldsymbol{\xi} \le \boldsymbol{w}^{\max}} \left \{  \boldsymbol{c}_{\alpha}^{\top} \boldsymbol{x} - \boldsymbol{c}_{\xi}^{\top} \boldsymbol{\xi} + \max_{\hat{ \boldsymbol{w}} \in \hat{ \mathcal{W} } }  \min_{\boldsymbol{y} \in \mathcal{Y}(\boldsymbol{x}, \boldsymbol{w}) ,\boldsymbol{w}, \boldsymbol{z}  } b^{\top} \boldsymbol{y} \right \}  \nonumber  \\
&&  \quad   \text{ s. t. } ~~ \quad    \boldsymbol{w} = \boldsymbol{z} \circ  \hat{\boldsymbol{w}} + (\boldsymbol{1} - \boldsymbol{z} )\circ \boldsymbol{\xi} \\
&& \qquad \qquad  ~~ \boldsymbol{w}  \le \boldsymbol{\xi}, ~~ \boldsymbol{w}  \le \hat{\boldsymbol{w}}    \nonumber 
\end{eqnarray}
in which \( (\boldsymbol{w}, \boldsymbol{z}) \) is part of the variables of the inner minimization problem. Actually, given \(( \boldsymbol{\xi},\hat{ \boldsymbol{w}}) \), there is only one feasible point \( (\boldsymbol{w}, \boldsymbol{z})\).
Therefore, the DDU-based RO is turned into a DIU-based RO problem, of which the second-stage problem is a mixed-integer SOCP. 
This problem with binary variables in the second stage can be solved by the nested C\&CG algorithm.
However, since the objective function \(  Q( \boldsymbol{x}, \boldsymbol{\xi}, \hat{\boldsymbol{w}} ):=\min b^\top \boldsymbol{y}\) in the maximization problem is not quasi-convex in $\hat{\boldsymbol{\omega}}$, the worst-case \(  \hat{\boldsymbol{w}}^{\ast} \) may reside in the interior of the uncertainty set \( \hat{ \mathcal{W} } \).
Therefore, the nested C\&CG algorithm cannot guarantee finite-step convergence.

\subsubsection{Modified Benders dual decomposition \cite{zhang2022two}}
The dual of \( \max_{\boldsymbol{w} \in \mathcal{W}(\boldsymbol{\xi}) }  (-\lambda_t^{\top }G\boldsymbol{w}) \) is considered after taking the dual of the inner minimization problem.
Using such double duality, advanced feasibility and optimality cuts to refine first-stage decisions are added to the master problem. 
Like Benders decomposition, this algorithm returns constraints with dual variables from the second stage.
Although its finite-step convergence for solving linear RO has been proven, it is not for nonlinear problems.
Applying it to RDNR \eqref{P：DNR} leads to the following subproblem:
\begin{eqnarray}\label{bender_dual}
&& \max_{ \boldsymbol{\lambda} \ge 0 ,\boldsymbol{\mu} }   ~ (A  \boldsymbol{x} + \gamma )^{\top } \lambda_{t}+ D( \lambda_t )   \nonumber  \\
&&    \text{ s. t. } ~~   \boldsymbol{b} + B^{\top }\lambda_t  =  C^{\top }\boldsymbol{\mu}+ D^{\top } \lambda_h    \\
&&  ~~  \quad   \quad \left \| \mu _{ij,t} \right \|_2 \le \lambda _{h,ij,t}, ~ \forall (i,j) \in \mathcal{E}, ~ t \in  \mathcal{T}   \nonumber
\end{eqnarray}
where
\begin{eqnarray}
D( \lambda_t ) &:=& \min_{\boldsymbol{\pi} \ge 0, \boldsymbol{\theta} \ge 0}   f^{\top }\boldsymbol{\pi} + \boldsymbol{\xi}^{\top }\boldsymbol{\theta}
\nonumber\\
\textnormal{s. t.} &&  F^{\top }  \boldsymbol{\pi} + \boldsymbol{\theta} + G^{\top }  \lambda_t  = 0   \nonumber
\end{eqnarray}
Problem \eqref{bender_dual} is an SOCP, whose feasible set of \( (\boldsymbol{\lambda}, \boldsymbol{\mu})\) is generally not a polyhedron, which may lead to an infinite number of extreme points and the failure of finite-step convergence.

\subsubsection{Mapping-based C\&CG (proposed algorithm)}
A mapping from the first-stage decision to a potential worst-case scenario is generated and returned with columns and constraints to the master problem.
The mapping is constructed by the KKT condition to indicate an extreme point of the uncertainty set \(  \mathcal{W}(\boldsymbol{\xi}) \) for given \( \boldsymbol{\xi}  \).
Therefore, the proposed algorithm applied to the RDNR problem can converge to the optimal solution in finite steps by enumerating the finite number of extreme points.
In case studies, we compare the proposed algorithm with the modified Benders dual decomposition, and the results show that our algorithm converges in finite steps with greatly reduced computation time.
Moreover, the proposed method can be easily extended to three-phase networks (which extend SOCPs in single-phase networks to semidefinite programs \cite{gan2014convex}), as well as multi-stage RO problems with DDU.

\section{Sensitivity to Uncertainty Set Parameters}\label{section4}
The uncertainty sets in RO are often constructed using rich historical data and advanced forecasting techniques \cite{bertsimas2009constructing}.
However, the parameters of uncertainty sets, which control the critical trade-off between security and efficiency (less conservativeness), are not always accurate, due to limited accuracy of prediction.
In \cite{xie2024robust}, an uncertainty set is modeled based on paid predictions from independent agents, where a higher prediction accuracy requires higher payments.
In this context, analyzing the sensitivity of the optimal value of RO with respect to the uncertainty set parameters is essential to deciding a reasonable prediction accuracy (and cost) when constructing the uncertainty set.

To analyze such a sensitivity for the RDNR problem considered, we first study the effect of changing uncertainty set parameters \(f\) in \eqref{dual_uncertainty_set} on the second-stage optimal objective value in \eqref{max_w}. Note that
\(\lambda_t^{\ast}\) in \eqref{max_w} is the optimal solution of \eqref{first-dual} given \( \left (   \boldsymbol{x}^{\ast },  \boldsymbol{\xi}^{\ast }   \right )  \).
The dual problem of \eqref{max_w} is:
\begin{eqnarray}\label{dualmax_w}
&&  \min_{\boldsymbol{\pi} \ge 0, \boldsymbol{\theta} \ge 0} ~~   f^{\top } \boldsymbol{\pi} + \boldsymbol{\xi}^{\top }  \boldsymbol{\theta}  \nonumber \\
&&    \text{ s. t. } ~~  G^{\top } \lambda_t^{\ast}  + F^{\top } \boldsymbol{\pi} +  \boldsymbol{\theta} = 0  
\end{eqnarray}
where the optimal \((\boldsymbol{\pi}, \boldsymbol{\theta} )\) can be interpreted as the sensitivity of the second-stage optimal value with respect to the uncertainty set parameters.

Further, the first-stage solution of RDNR consists of the network topology and the RG resizing.
Denote their sensitivities by \( \boldsymbol{\vartheta}_{x} \) and \( \boldsymbol{\vartheta}_{\xi} \), respectively.
The optimal resizing solution $\boldsymbol{\xi}^{\ast}$ determines the upper bound of the uncertainty set as \( \boldsymbol{w} \le \boldsymbol{\xi}^{\ast}\), so its sensitivity \( \boldsymbol{\vartheta}_{\xi}\) is the smaller of the optimal \( \boldsymbol{\theta} \) in \eqref{dualmax_w} and the cost factor \( \boldsymbol{c}_{\xi}\) in \eqref{P：DNR}.

\begin{figure}
	\centering
\includegraphics[width=0.65\columnwidth]{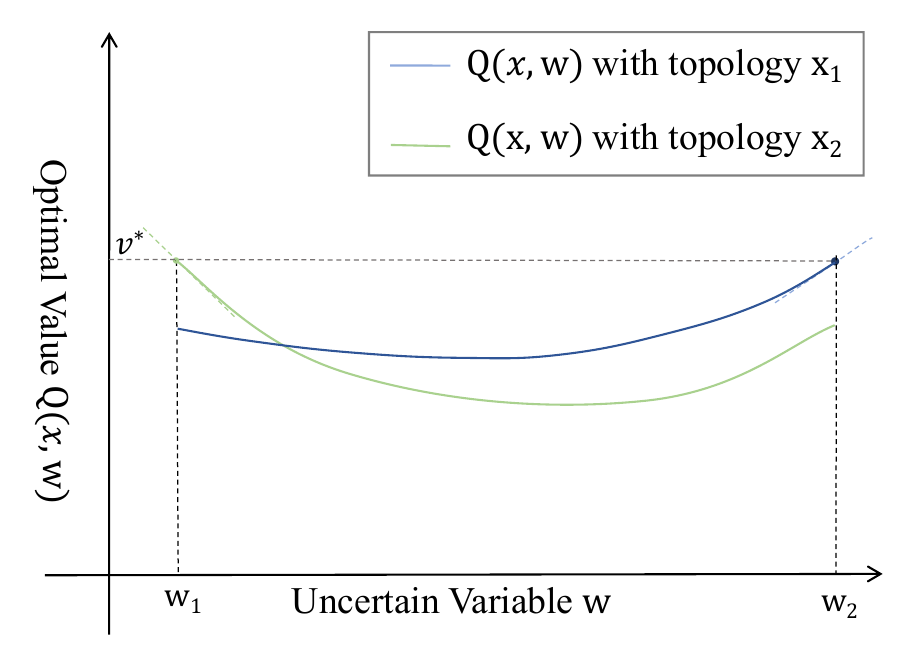}
	\caption{The second-stage optimal value $Q(\boldsymbol{x}, \boldsymbol{w})$ reaches the same worst-case value under topologies $\boldsymbol{x}_1, \boldsymbol{x}_2$.}	
 \label{section_4}
\end{figure}

We next discuss the influence of the uncertainty set on the optimal topology $\boldsymbol{x}^{\ast}$. 
Note that for each given topology $\boldsymbol{x}$, the second-stage optimal value $ Q\left (   \boldsymbol{x},  \boldsymbol{w} \right )$ is a continuous function in $\boldsymbol{w}$, on which a worst-case (maximizing) $\boldsymbol{w}^{\ast}(\boldsymbol{x})$ is attained at an extreme point of the uncertainty set. 
If there are more than one optimal topologies such as the $\boldsymbol{x}_1$ and $\boldsymbol{x}_2$ shown in Fig. \ref{section_4}, which reach the same (minimum) $Q\left (\boldsymbol{x}_1,  \boldsymbol{w}^{\ast}(\boldsymbol{x}_1)\right ) = Q\left (\boldsymbol{x}_2,  \boldsymbol{w}^{\ast}(\boldsymbol{x}_2)\right) = v^{\ast}$, then perturbing the uncertainty set by a small amount might change the optimal topology. 
For instance, if the uncertainty set is reduced by a small number $\epsilon$ into $[w_1+\epsilon,~w_2]$, then $\boldsymbol{x}_2$ would be the only optimal topology, while $\boldsymbol{x}_1$ is no longer optimal. 
Other than such special cases (of more than one optimal topologies), in general, a sufficiently small change in the uncertainty set would not alter the optimal topology.    
In practice, a solver may return only one optimal topology even if multiple ones exist. To address this challenge, we sample small variations in the uncertainty set parameters to identify all possible changes in optimal topologies. Then, based on the optimal solution of \eqref{dualmax_w} with $\lambda_t^{\ast}$ taken under different topologies, we can derive the sensitivity of the optimal topology with respect to the uncertainty set parameters.

\section{Case Study}\label{case_study}

We experiment with several IEEE networks under different forms of uncertainty sets and different numbers of time periods to examine the performance of the proposed algorithm.
All experiments are run in MATLAB with the optimization solver GUROBI, on a laptop computer with an AMD Ryzen 7, 6800H, 3.20GHz CPU and 16GB RAM.

\subsection{Experimental Setup}

\begin{table}
\caption{Information of test networks}
\begin{tabular}{cccc}
\hline 
\text { Test network } & \text { 13-bus } & \text { 33-bus } & \text { 123-bus } ~~ \\
\hline \text { No. of RG buses } & 3 & 6 & 6 ~~~ \\
\text { No. of BES buses} & 2 & 4 & 8 ~~~ \\
\text { No. of redundant branches} & 3 & 5 & 5 ~~~ \\
\hline
\end{tabular}\label{tab:table1}
\end{table}

\begin{figure}
	\centering
\includegraphics[width=0.9\columnwidth]{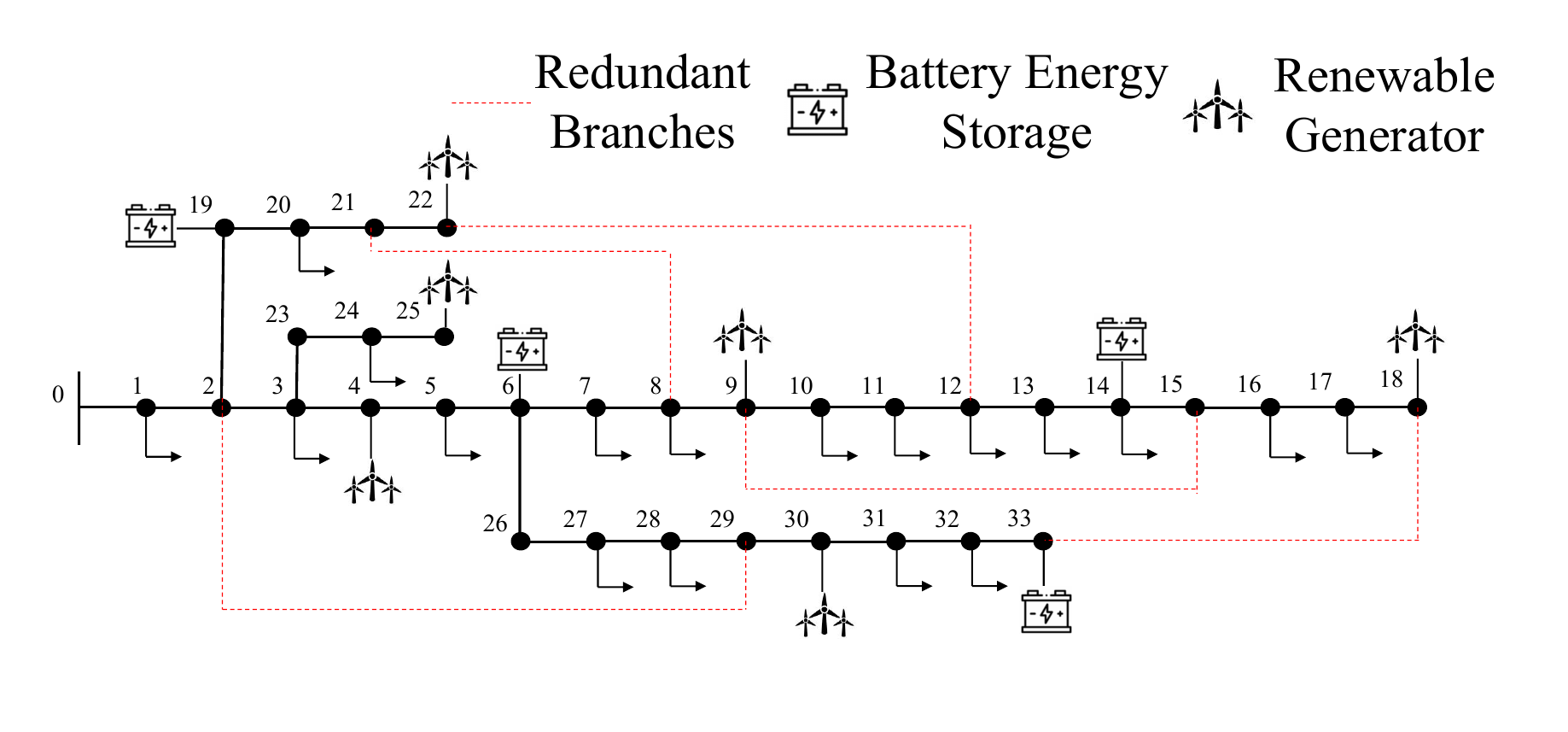}
	\caption{IEEE 33-bus distribution network with five redundant branches.}	
 \label{33-bus}
\end{figure}

We simulate IEEE 13, 33, 
and 123-bus distribution networks with their information summarized in
Table \ref{tab:table1}. The 33-bus network is also illustrated in Fig. \ref{33-bus} as an example.

\begin{figure}
	\centering
\includegraphics[width=0.8\columnwidth]{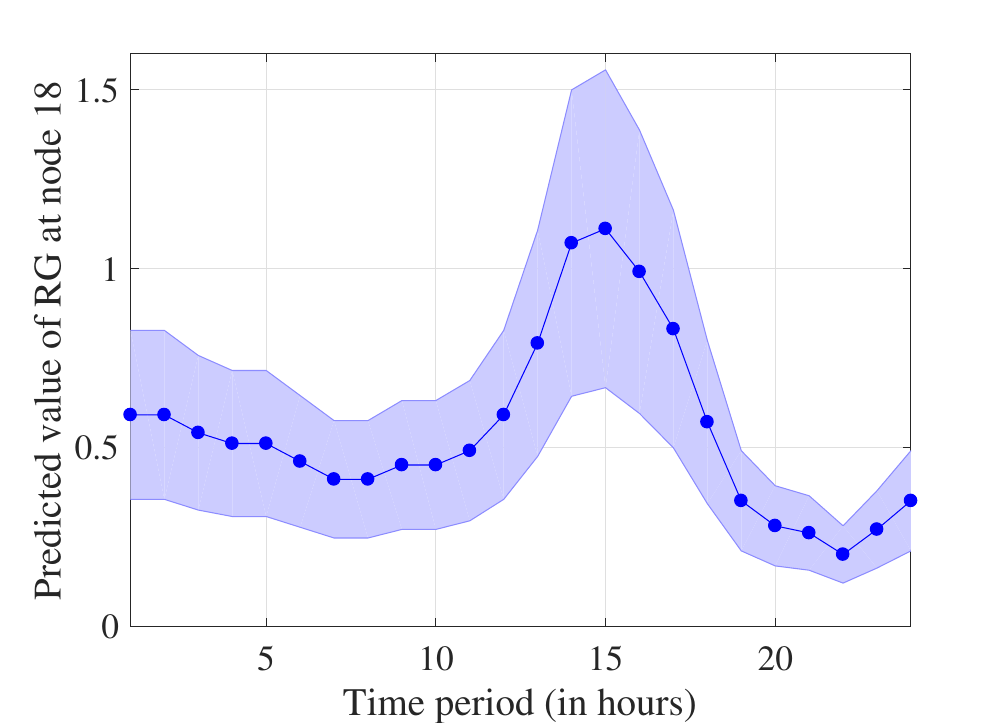}
	\caption{Prediction and range of RG output at bus 18, IEEE 33-bus network.}	
 \label{predict-data}
\end{figure}

We assume the switching cost is uniform across all lines to simplify the computation, although in practice the switching costs can be different depending on the aging, hardening, or operating risk of different lines.
The load and RG output predictions are based on the data from Germany in October 2020 \cite{OPSDP2022}, with 15-minute resolution (for experiments with $T=4$ time periods) or hourly resolution (for experiments with $T=12$ and $24$ time periods).
These data are scaled to match the test network capacities.
The ranges \([w_{i,t}^{\textnormal{min}} ,~w_{i,t}^{\textnormal{max}}]\) of RG outputs are set at 0.5 and 1.5 times their predicted outputs \(w_{i,t}^{p}\).
The spatial and temporal deviation budgets are set as \( \Gamma_i = 0.5  T \) (only for $T>1$ time periods) and \( \Gamma_t = 0.5  N_w\).
Fig. \ref{predict-data} illustrates the prediction and range of RG output at bus 18 of the IEEE 33-bus network.

\subsection{Convergence Performance}

\begin{figure}
	\centering
\subfigure[]{
\includegraphics[width=0.8\columnwidth]{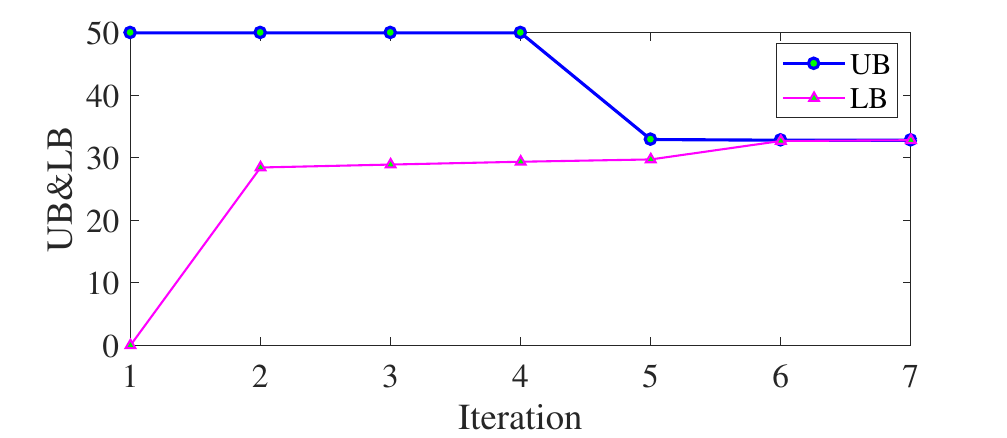}
}\label{without}
\subfigure[]{
\includegraphics[width=0.8\columnwidth]{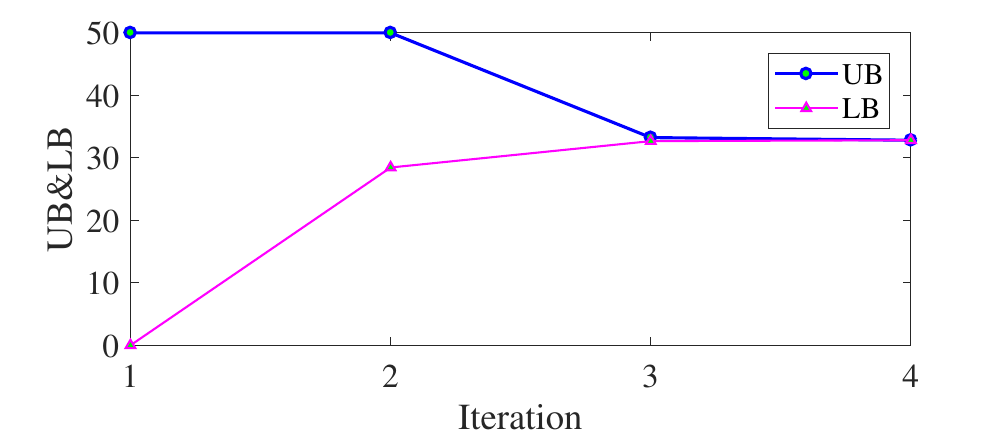}
}\label{with}
	\caption{Convergence of the mapping-based C\&CG algorithm (Algorithm \ref{alg:DualCCG}) in the IEEE 33-bus network. Subfigure (a) adds feasibility cuts \cite{chen2022robust, zhang2022two, zeng2022two} when $(\boldsymbol{x},\boldsymbol{\xi})$ is not robustly feasible; (b) uses the relaxed formulation \eqref{relaxed-2}.}	
 \label{with_without}
\end{figure}

We first run the proposed algorithm at a single time period  on the IEEE 33-bus network, to verify its convergence behavior and computational efficiency.

Fig. \ref{with_without} shows the iteration of lower bound (LB) and upper bound (UB) in the proposed Algorithm \ref{alg:DualCCG}, under different choices to ensure feasibility of the second-stage problem: (a) adding feasibility cuts \cite{chen2022robust, zhang2022two, zeng2022two}, or (b) using the relaxed formulation \eqref{relaxed-2} and its dual, the \textbf{SP} \eqref{subproblem}. Under both choices, the proposed algorithm converges in a few steps. With the relaxed formulation \eqref{relaxed-2}, the convergence takes even fewer steps and shorter computation time than the traditional feasibility cut method: (b) 3.94 seconds versus (a) 8.59 seconds.  

\begin{figure}
	\centering
\includegraphics[width=0.8\columnwidth]{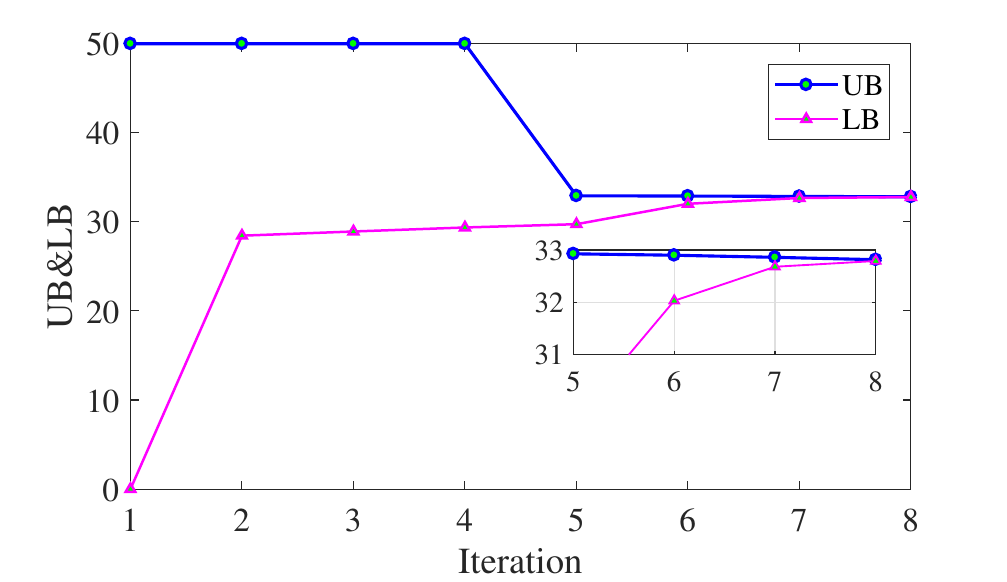}
	\caption{Iteration of the modified Benders dual decomposition algorithm in the IEEE 33-bus network.}	
 \label{bender-dual}
\end{figure}

\begin{table}
\caption{Comparison of computing efforts between the proposed algorithm and modified Benders decomposition.}
\centering
\begin{tabular}{l l c c}
\toprule
\multirow{2}{*}{Test network} & \multirow{2}{*}{} & \multicolumn{2}{c}{Algorithm} \\
\cmidrule(lr){3-4}
& & Proposed algorithm & Modified Benders \\
\midrule
\multirow{2}{*}{13-bus} 
& Time (s)     & 1.79  & 3.81 \\
& Iterations  & 3     & 7 \\
\midrule
\multirow{2}{*}{33-bus} 
& Time (s)     & 3.94  & 9.72 \\
& Iterations  & 3     & 8 \\
\midrule
\multirow{2}{*}{123-bus} 
& Time (s)     & 10.04 & 41.75 \\
& Iterations  & 3     & 10 \\
\bottomrule
\end{tabular}\label{tab:compare}
\end{table}

For comparison, Fig. \ref{bender-dual} shows the iteration of the modified Benders dual decomposition algorithm \cite{zhang2022two} when solving the same RDNR problem as in Fig. \ref{with_without}. 
The gap (UB $-$ LB) converges to zero asymptotically, rather than in finite steps, due to the SOC constraints in the maximization problem.
The bilinear term \( \boldsymbol{\xi}^{\top }\boldsymbol{\theta}  \) added to the constraints of the master problem further increases the computational complexity.
Consequently, the time required for each iteration of the modified Benders dual decomposition algorithm grows with the number of iterations and exceeds that of the mapping-based C\&CG.
Table \ref{tab:compare} compares the computation time and the number of iterations of the proposed algorithm and the modified Benders algorithm, in different test networks, which verifies the improvement of the proposed algorithm in computational efficiency.

\begin{figure}
	\centering
\includegraphics[width=0.85\columnwidth]{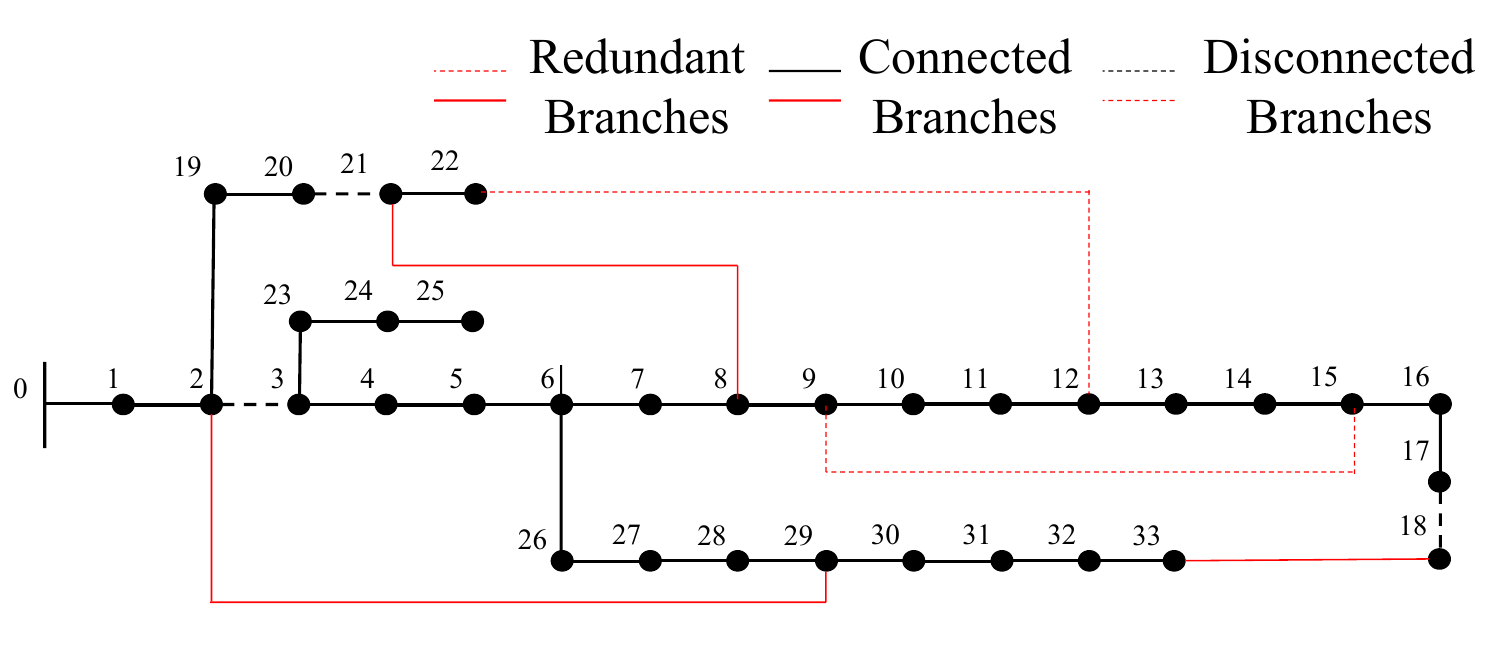}
	\caption{Optimal topology of RDNR in the IEEE 33-bus network.}	
 \label{optimal-topology}
\end{figure}

\begin{table}
\caption{Sensitivity of optimal value to uncertainty set parameters}
\centering
\begin{tabular}{lcc|c}
\toprule
\textbf{Bus index} & $w_{i}^{\text{min}}$ & $w_{i}^{\text{max}}$ & $\Gamma$ \\
\midrule
Bus 4  & 0     & 0     & \multirow{6}{*}{0.336} \\
Bus 9  & 0     & 0     &                     \\
Bus 18 & 0     & 1.42  &                     \\
Bus 22 & 0     & 1.57  &                     \\
Bus 25 & 0     & 0     &                     \\
Bus 30 & 0     & 1.08  &                     \\
\bottomrule
\end{tabular}
\label{tab:sensitivity}
\end{table}

We also analyze the impact of uncertainty set parameters \(  w_{i}^{\text{min}} \), \(w_{i}^{\text{max}}\) and \(   \Gamma\) (time index $t$ is skipped for the single time period problem) in the 33-bus network.
The unique optimal topology is shown in Fig. \ref{optimal-topology}.
The worst-case scenario under the optimal topology is \(  \boldsymbol{w}^{\ast} = [ 0.392,~0.498,~0.385,~0.402,~0.367,~ 0.280]  \), which is uniquely determined by the following active constraints:
\begin{subequations}
\begin{eqnarray}\label{active}
w_{18} =  w_{18}^{\text{min}}, ~~  w_{22} =  w_{22}^{\text{min}}, ~~  w_{30} =  w_{30}^{\text{min}},  \label{activea} \\
\frac{\left | w_{4}- 0.392 \right |  }{0.196}  + \frac{\left | w_{9} - 0.498 \right |  }{0.299} + \frac{\left | w_{25} - 0.367 \right |  }{0.1835}  = 0    \label{activeb}
\end{eqnarray}
\end{subequations}
where \eqref{activeb} is equivalently transformed from some constraints in \(  F \boldsymbol{w}  \le f  \). 
The sensitivity of the optimal value to the uncertainty set parameters, as computed by \eqref{dualmax_w}, is shown in Table \ref{tab:sensitivity}.
When the lower bounds of \( w_{18}, w_{22} \) and \( w_{30} \) make small changes, the optimal value will change accordingly.
The positive sensitivity numbers indicate that the optimal cost decreases with the contraction of the uncertainty set.

\subsection{Performance on Different Uncertainty Set}

\begin{figure}
	\centering
\includegraphics[width=0.8\columnwidth]{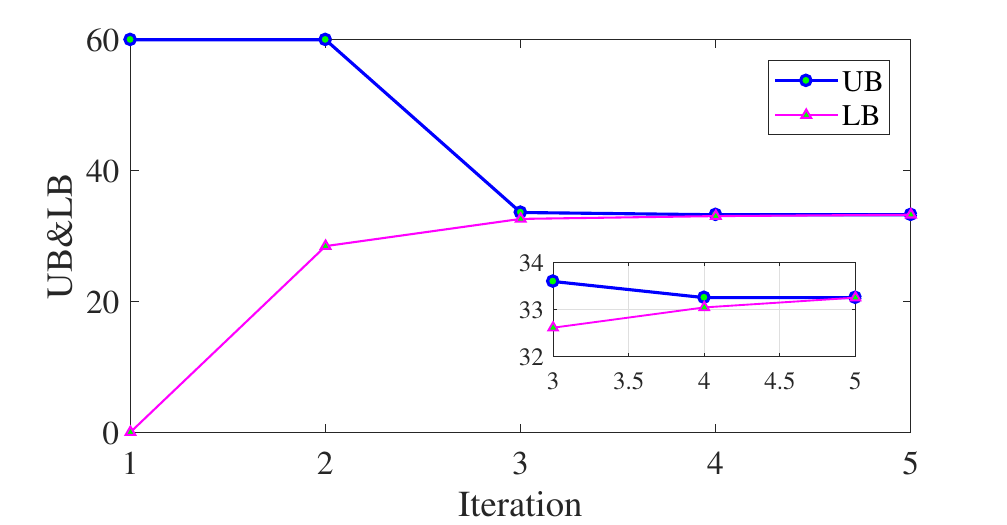}
	\caption{Convergence of the proposed algorithm with an added constraint \eqref{add_con} to the uncertainty set.}	
 \label{add}
\end{figure}

To show good applicability of the proposed algorithm, we consider a different form of uncertainty set for the single time period RDNR problem in the 33-bus network.
We add the following constraint to the uncertainty set:
\begin{eqnarray}\label{add_con}
&&  \sum_{j \in \mathcal{N}_w, j \neq i}  w_j  \ge - \xi_{i} +  \beta_{i} , \quad \forall i \in \mathcal{N}_w.
\end{eqnarray}
The added constraint \eqref{add_con} provides a supporting lower bound with the first-stage variable \( \boldsymbol{\xi} \), which makes it decision-dependent.
We rewrite \eqref{add_con} in a compact form:
\begin{eqnarray}\nonumber
&&  H \boldsymbol{w}   \ge \boldsymbol{\beta}    - \boldsymbol{\xi}  : \boldsymbol{\sigma}
\end{eqnarray}
which changes the KKT condition \eqref{kkt-condition} accordingly. The RDNR problem is still solvable by the proposed algorithm in finite steps, as shown in Fig. \ref{add}.

\subsection{Scalability}

\begin{table}
\caption{Computing efforts of the proposed algorithm in different networks and numbers of time periods}
\begin{tabular}{ccccc}
\hline 
\text { No. of time periods $T$ } & $1$ & $4$ & $12$ & $24$ \\
\text { Test network } &     \multicolumn{4}{c}{Computing time in seconds (Iterations)}                    
\\
\hline \text { 13-bus }
      & 1.79 (3)  &   107.9 (3)  &     422 (3)  & 1688 (3) \\
 \text { 33-bus} 
   & 3.94 (3) &  150.7 (3) &   834 (3)  &  3712 (4)  \\
\text { 123-bus} 
   & 10.04 (3)  &   293.1 (3)  &     3583 (3)  &  -  \\
\hline
\end{tabular}\label{tab:table2}
\end{table}

Table \ref{tab:table2} shows the computing time and the number of iterations for the proposed algorithm to solve RDNR with different time periods in different networks.
As the network grows, the computing time increases moderately.
The increase of computational burden with the number of time periods is more significant, partly due to the drastically increasing number of uncertainty variables, which is acceptable for the day-ahead network reconfiguration.

\section{Conclusion}\label{section6}

We presented a robust optimization model for distribution network reconfiguration (DNR), which incorporates renewable generator (RG) resizing with network topology optimization and introduces decision-dependent uncertainty. 
We developed a mapping-based C\&CG algorithm to solve the problem formulated and showed that it can converge to an optimal solution in finite steps.
Sensitivity of the optimal solution to uncertainty set parameters was analyzed.
The finite-step convergence and computational efficiency of the proposed algorithm have been verified by case studies. 
This work provides a robust framework for improving power system reliability and efficiency in the presence of increasing renewable energy penetration.

Looking forward, this work may be extended in several directions. 
First, the proposed framework can be adapted to unbalanced three-phase distribution networks by utilizing semidefinite relaxation techniques. 
Second, we will consider a multi-stage dynamic DNR problem with RG resizing and energy storage management, to use renewable energy more efficiently.
We will also incorporate renewable generation and load forecasting decisions into the construction of decision-dependent uncertainty sets.

\bibliographystyle{IEEEtran}
\bibliography{references}

\end{document}